\documentclass[10pt,conference]{IEEEtran}

\AtBeginDocument{
  }

\usepackage{cite}

\usepackage{float}
\usepackage{balance}
\usepackage{graphicx}

\usepackage[dvipsnames]{xcolor}
\usepackage{pifont}
\usepackage{booktabs}
\usepackage{amsmath}
\usepackage{mathrsfs}

\makeatletter
\let\Bbbk\@undefined
\makeatother
\usepackage{extpfeil}

\usepackage{tcolorbox}
\newtcolorbox{mybox}{colback=black!5!white,colframe=black,bottomrule=.3mm,toprule=.3mm,leftrule=.3mm,rightrule=.3mm,left=.3mm,right=.3mm,top=-.3mm,bottom=-.3mm}

\usepackage{subfigure}

\usepackage{listings}
\lstdefinelanguage{Tamarin}
{
  morekeywords={lemma, All, Ex, not, exists, trace},
  keywordstyle=\color{blue}\bfseries,
  sensitive=true,
  morecomment=[l]{//},
  commentstyle=\color{gray}\ttfamily,
  morestring=[b]{},
  stringstyle=\color{black},
}

\lstdefinestyle{customtamarin}{
  language=Tamarin,
  basicstyle=\ttfamily,
  keywordstyle=\bfseries,
  stringstyle=\color{black},
  commentstyle=\color{gray}\itshape,
  breaklines=true,
  showstringspaces=false,
}

\newcommand{\cmark}{\ding{51}}
\newcommand{\xmark}{\ding{55}}

\colorlet{darkergreen}{green!50!black}
\definecolor{darkred}{rgb}{0.6, 0.0, 0.0}

\newcommand{\falsified}{\textcolor{darkred}{\xmark}}
\newcommand{\verified}{\textcolor{darkergreen}{\cmark}}

\newcommand{\seqa}{\(A2B\)}
\newcommand{\seqb}{\(B2A\)}

\newcommand{\bhd}[1]{\noindent  \textbf{#1.}\quad}

\newcommand{\ihd}[1]{\noindent  \textit{#1.}\quad}

\usepackage{tabularx}
\usepackage{booktabs}
\usepackage{array}

\usepackage{enumitem}

\definecolor{keywordcolor}{rgb}{0.8, 0.0, 0.0}
\definecolor{paramcolor}{RGB}{128,0,0}
\definecolor{textcolor}{RGB}{0,0,0}

\usepackage{tikz}

\usetikzlibrary{arrows.meta, positioning}

\usepackage[hidelinks]{hyperref}

\newcommand{\circled}[1]{\tikz[baseline=(char.base)]{
  \node[shape=circle,draw,inner sep=0.5pt] (char) {#1};}}

\begin{document}

\title{Beyond the Quantum Promise: A Security Analysis of Classical Control in Quantum Key Distribution}

\author{
\IEEEauthorblockN{Ali Hamza Malik}
\IEEEauthorblockA{University of Massachusetts Amherst\\
Amherst, MA, USA\\
ahmalik@umass.edu}
\and
\IEEEauthorblockN{Raja Hasnain Anwar}
\IEEEauthorblockA{University of Massachusetts Amherst\\
Amherst, MA, USA\\
ranwar@umass.edu}
\and
\IEEEauthorblockN{Muhammad Taqi Raza}
\IEEEauthorblockA{University of Massachusetts Amherst\\
Amherst, MA, USA\\
taqi@umass.edu}
}

\maketitle

\begin{abstract}

Quantum Key Distribution (QKD) protocols provide information-theoretic security by using quantum mechanical principles. Yet QKD is fundamentally a \emph{hybrid} protocol: its security depends on the correct integration of the quantum phase with classical post-processing. While ETSI and ITU-T specifications standardize QKD architectures and interfaces, they evaluate protocol security in isolation, leaving cross-layer interactions as an underexplored attack surface.

This paper introduces a formal verification framework that holistically models QKD protocols based on ETSI and ITU-T QKD specifications. Our model is the first hybrid QKD protocol model that supports automated analysis of protocol-level security focusing on how classical operations influence the security guarantees provided by the quantum phase of the QKD protocol. We formalize a comprehensive symbolic model of QKD protocols, based on ETSI and ITU-T QKD specifications, in Tamarin, an automated protocol verifier.

Applying this framework, we obtain formal evidence of three specification-level vulnerabilities in ETSI- and ITU-T-grounded protocol models under adversary \(Eve^+\): subverted entanglement injection, basis-deferred measurement, and message reflection. Each arises from a classical control-plane omission in the procedure text and is established under a symbolic abstraction rather than as a claim about all practical deployments. We introduce two protocol improvements: measurement commitment and identity-bound message authentication codes (MACs). Tamarin verification confirms that these countermeasures eliminate the identified vulnerabilities under \(Eve^+\). We have communicated our results and recommendations to relevant standardization organizations.

\end{abstract}

\section{Introduction}

\begin{figure}[t]
    \centering
    \includegraphics[width=0.95\linewidth]{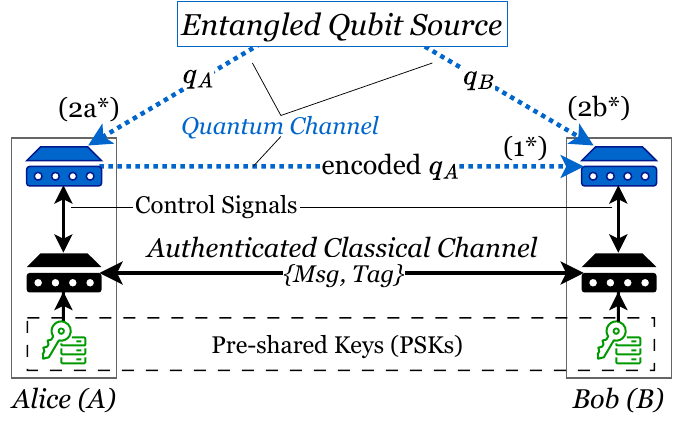}
    \vspace{-5pt}
    \caption{A typical two-node QKD architecture:
   \texttt{(1*)} shows a one-way quantum channel from Alice to Bob for PM QKD. (\texttt{2a*} and \texttt{2b*}) depict quantum channels from a shared source of entangled qubit pairs to Alice and Bob for EB QKD. The \textit{control signals} illustrate the interaction between quantum and classical phases.}
   \label{fig:Arch_QKD}
   \vspace{-10pt}
\end{figure}

Quantum Key Distribution (QKD) is among the most mature deployed applications of quantum networking, with commercial links spanning metropolitan fiber and satellite channels~\cite{cao2022evolution,qkdnetwork,mehic2023quantum,Pirandola_2020,rfc9340}. Unlike computationally secure key exchange, QKD targets information-theoretic security~\cite{shor2000simple,portmann2014cryptographicsecurityquantumkey,tomamichel2017largely,gisin2002quantum,etsi_qkd_005} through quantum limits such as the no-cloning theorem~\cite{buvzek1996quantum} and measurement uncertainty~\cite{xiao2023quantum}. QKD is inherently \emph{hybrid}: a quantum phase establishes raw key material, and a classical phase executes sifting, error estimation, error correction, and privacy amplification (Figure~\ref{fig:Arch_QKD}). ETSI and ITU-T specifications standardize architectures and interfaces, yet security arguments for QKD frequently treat individual layers or phases in isolation rather than as one integrated protocol.

Prior work on QKD security focuses primarily on quantum-layer attacks, such as intercept-resend and photon-number-splitting~\cite{wiesemann2024evaluation,sixto2024quantum,xu2024automatically}. Classical post-processing steps, specifically sifting, reconciliation, and privacy amplification, have received comparatively less formal scrutiny as an integrated attack surface. Error-correction analyses~\cite{pfister2016sifting,tomamichel2017largely} typically assume ideal classical channels and do not model leakage or forgery scenarios that couple to quantum measurement outcomes. No prior analysis spans both prepare-and-measure (PM) and entanglement-based (EB) variants with authenticated classical channels under a unified symbolic model. Cross-layer attack patterns, such as replayed basis reconciliation and deferred sifting, exemplify the threat class that a strictly quantum or strictly classical model cannot capture, and motivate the hybrid analysis presented here.

This work presents a standards-grounded framework for verifying QKD session properties across quantum and classical phases jointly. Three constraints shape the approach. \emph{Informal standards} complicate precise models of procedure order and dependencies~\cite{kuppam2016modelling,nagarajan2005automated,nagarajan2002formal,fernandez2011formal,kubota2011formal}. \emph{Stateful sessions} are those in which later steps depend on earlier quantum observations and accumulated transcripts. \emph{Probabilistic quantum operations} must be abstracted for tractable automated reasoning~\cite{ying2019model,hirschi2019symbolic}.

We interpret ETSI and ITU documents (Appendix, Table~\ref{tab:qkd_standards}) to extract procedures and variants, enumerate underspecified choices, and construct symbolic protocol models. We define the threat model and verification pipeline, comprising symbolic quantum encoding, security lemmas, and the Tamarin toolchain, in \S\ref{sec:methodology_section}. We abstract probabilistic measurement as property-driven symbolic functions, extend the Dolev--Yao model with quantum knowledge to cover hybrid attack surfaces, and target \emph{session-key secrecy} and \emph{entity authentication}. Formal verification relies on the Tamarin Prover~\cite{meier2013tamarin,basin2022tamarin}. Lemmas, templates, and counterexample traces appear in \S\ref{sec:methodology_section}.

Across nine protocol configurations under adversary \(Eve^+\), we evaluate four security checks per configuration and identify three specification-level vulnerability classes that break secrecy or authentication in the symbolic model (Section~\ref{section:traceanalysis}). We derive two protocol-level countermeasures, re-verify them in Tamarin, and report the findings in Section~\ref{sec:discussion_section}. \textbf{QVerify}\footnote{\url{https://github.com/KhwarizmiLab/QVerify}}, described in \S\ref{sec:methodology_section} and Appendix~\ref{appendix:artifact}, packages the models and verification scripts.

\bhd{Contributions}
\begin{itemize}
    \item \textbf{Standards-grounded models.} We derive protocol assumptions, security goals, and procedure variants from ETSI and ITU-T sources (Appendix~\ref{appendix:tab}).

    \item \textbf{Hybrid symbolic verification.} We construct a unified Tamarin model family for PM and EB QKD with symbolic quantum operations and adversary \(Eve^+\), as defined in \S\ref{sec:methodology_section}. Prior symbolic abstractions provided foundational methods for isolated quantum properties~\cite{hirschi2019symbolic}. Our unified model jointly encodes quantum and classical phases and exposes cross-layer attack patterns that arise from their interaction. \textbf{QVerify} is released with full source code, including Tamarin models, \texttt{m4} templates, and verification scripts~\cite{QVerify2025}.

    \item \textbf{Verification at scale.} We instantiate nine target models from a single \texttt{m4} template and report secrecy and agreement lemma results across all configurations (Table~\ref{tab:results_group}).

    \item \textbf{Vulnerabilities and fixes.} Under adversary \(Eve^+\) in the symbolic model, we identify three specification-level vulnerability classes rooted in classical control-plane omissions---subverted entanglement injection, basis-deferred measurement, and message reflection (Section~\ref{section:traceanalysis})---propose two countermeasures (Section~\ref{sec:countermeasures}), and Tamarin verification confirms that the combined fix set restores the target properties under \(Eve^+\).
\end{itemize}

\section{Background on QKD Protocol}
\label{sec:qkdnetworkarchitecture}

\subsection{QKD Network Architecture}

A typical QKD deployment connects Alice and Bob with an insecure quantum channel and an authenticated classical channel (Figure~\ref{fig:Arch_QKD}). Public device identifiers mark endpoints but do not authenticate users. Parties authenticate classical traffic with a pre-shared symmetric key (PSK), e.g., \(pskAB\). We assume PSKs are delivered securely. Scalable PSK distribution remains an open problem~\cite{nikolopoulos2024quantum,wang2021experimental}.

\subsection{QKD Protocols}
\begin{figure}
     \centering
     \includegraphics[width=\linewidth]{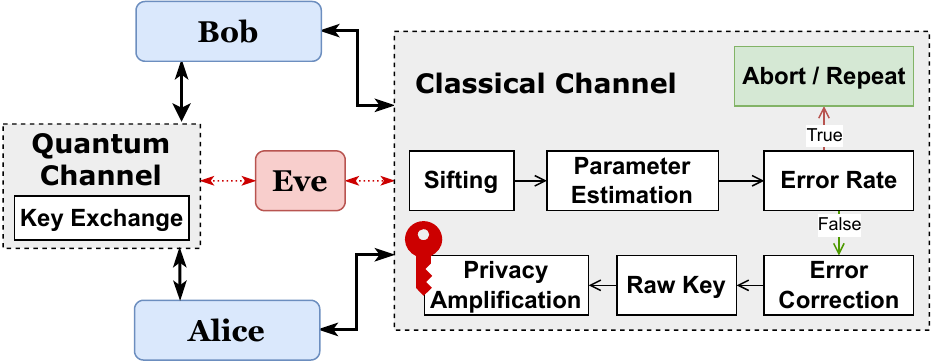}
    \caption{Overview of a typical QKD protocol. Alice and Bob exchange quantum states over an insecure quantum channel and communicate over an authenticated classical channel.
    }
     \label{fig:Overview_QKD}
 \vspace{-10pt}
\end{figure}

QKD splits into prepare-and-measure (PM)~\cite{bennett2014quantum,gisin2002quantum,shor2000simple} and entanglement-based (EB)~\cite{ekert1991quantum,zhong2022realistic,neumann2022experimental} families. Figure~\ref{fig:Overview_QKD} summarizes the shared structure: quantum exchange, then classical sifting, parameter estimation, reconciliation, and privacy amplification.

\subsection{Prepare-and-Measure QKD Protocols}
In PM QKD, Alice prepares qubits from random bits and bases, sends them to Bob, and Bob measures in random bases. After transmission, the parties publish which bases they used (not the bit values), discard mismatches, and keep the sifted bits. Parameter estimation samples sifted bits to estimate the quantum bit error rate (QBER) and detect eavesdropping. If QBER is too high, the session aborts (e.g., BB84 tolerates a QBER below 11\% under standard security analyses~\cite{shu2023asymptotically,shor2000simple,kim2008implementation,chong2010quantum}). ETSI and ITU-T specifications leave the ordering of basis announcement underspecified. For example, \cite{itu_FG_QIT4N_D2_3_1} describes the quantum layer protocol steps without mandating a sequence constraint between Alice's basis disclosure and Bob's qubit measurement. As Section~\ref{section:traceanalysis} shows, announcing Alice's basis before Bob measures enables basis-deferred measurement when the implementation follows that order. Error reconciliation and privacy amplification then run over the authenticated classical channel.

\subsection{Entanglement-Based QKD Protocols}

In EB QKD, a source distributes entangled pairs to Alice and Bob. We focus on an untrusted external source~\cite{ent_middle1,ent_middle2}. Each party measures in random bases, then performs sifting as in PM. For parameter estimation, mismatched-basis outcomes support CHSH tests~\cite{clauser1969proposed}: a Bell violation indicates shared entanglement, while classical bounds signal tampering or excess noise~\cite{gisin2002quantum}. EB security assumes correlations are evaluated on data that truly came from one distributed pair. Timing and synchronization between parties are therefore security relevant when classical steps are interleaved with quantum events.

\subsection{Formal Verification Background}
\bhd{Symbolic Protocol Verification} The symbolic model represents protocol messages as algebraic terms constructed from idealized cryptographic primitives~\cite{dolev1983security,stallings1995network}. The Dolev-Yao attacker~\cite{dolev1983security} has full network control and manipulates messages but cannot break the assumed perfect cryptography, as defined by the underlying equational theory. Security properties such as secrecy and authentication are checked by exhaustively analyzing reachable states under this model~\cite{cervesato2001dolev}.

\bhd{The Tamarin Prover} Tamarin Prover~\cite{meier2013tamarin,basin2022tamarin} automates symbolic security protocol verification. Protocols are described by labeled multiset rewrite rules. Security properties are stated as temporal logic lemmas over traces of rule actions. Tamarin supports automated and interactive proof search, and has been used to verify protocols including TLS~1.3~\cite{tamarintls}, 5G-AKA~\cite{basin2018formal}, EMV~\cite{basin2021emv}, Signal~\cite{cohn2020formal}, and iMessage PQ3~\cite{cryptoeprint:2024/1395,stebila2024security,basin2017symbolically}.

\section{Design and Analysis of Formal QKD Model}
\label{sec:methodology_section}

\begin{figure}[t]
    \centering
    \includegraphics[width=0.9\linewidth]{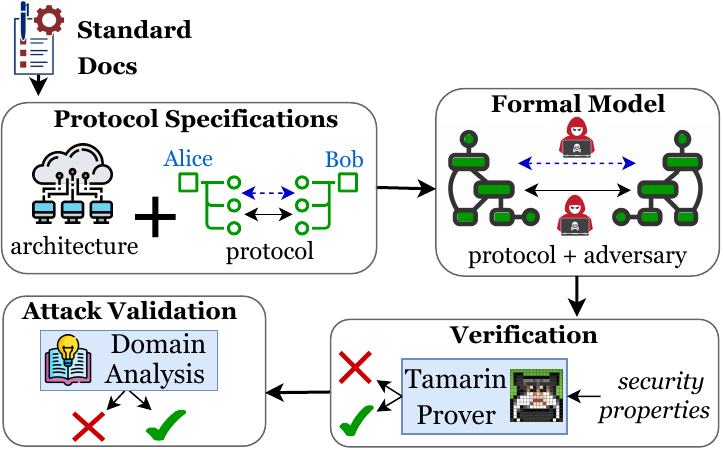}
    \vspace{-2pt}
    \caption{Overview of our methodology.}
    \label{fig:methodology}
    \vspace{-13pt}
\end{figure}

We build an abstract QKD model, grounded in ITU and ETSI procedures (Appendix~\ref{appendix:tab}), that ties classical messages to quantum observations for both PM and EB. We formalize channel layout, synchronization assumptions, and session procedures (basis exchange, reconciliation), then encode them for automated verification. Figure~\ref{fig:methodology} summarizes the pipeline.

\bhd{Threat Model}
\label{sec:threat_model}
We consider a quantum Dolev-Yao-style network adversary~\cite{hirschi2019symbolic,cervesato2001dolev} \(Eve^+\) with the capabilities of quantum eavesdropper \(Eve\) and classical network adversary \(Adv^C\), and we model quantum and classical cryptographic primitives as ideal in this analysis. \(Eve\), the traditional adversary in QKD security analysis, can store, measure, and inject qubits over the insecure quantum channel. \(Adv^C\) can drop, modify, or inject messages in the public classical channel. \(Eve^+\) cannot forge or alter authenticated classical messages without invalidating the MAC. The no-cloning theorem prevents \(Eve^+\) from copying an unknown qubit, but if \(Eve^+\) already knows the classical values used to prepare a qubit (e.g., by intercepting a basis announcement), she can re-prepare a new qubit with the same encoding and substitute the original.

\subsection{Abstract QKD Modeling}
\label{subsection:QKD_model}
This subsection covers the key abstractions and modeling choices for QKD protocol representation and highlights the challenge of modeling probabilistic quantum operations.

\subsubsection{Protocol Model}
\label{subsec:protocol_model}
We model the QKD protocol variants, i.e., \(P_{EB}\) (entanglement-based) and \(P_{PM}\) (prepare-and-measure), from the perspective of two participants: Alice \(A\) and Bob \(B\) in the presence of an eavesdropper \(Eve\), and we retain only the quantum exchange, sifting, error correction, and privacy amplification procedures, encoding the entanglement source and PSK distribution server in the \emph{environment} because their separate identities do not impact the analysis of relevant procedures. Throughout the model, the classical post-processing steps are defined as symbolic functions whose inputs are derived directly from prior quantum transmission and measurements, thereby capturing the interactions between quantum and classical phases.

We model the behavior of Alice \(A\) and Bob \(B\) as two labeled transition systems \(\mathcal{L}_A\) and \(\mathcal{L}_B\), respectively, which communicate with each other by transmitting and receiving messages and qubit batches. We implement the communication channels as separate unidirectional \textit{classical} and \textit{quantum} channels from \(\mathcal{L}_A\) to \(\mathcal{L}_B\) and vice versa, modeling the capabilities of eavesdropper \(Eve\) as properties of the quantum channels while qubit transmission determines the raw data and classical traffic supplies the operational interpretation of those quantum results (e.g., basis matching, syndrome exchange). We use unidirectional rather than bidirectional channels to simplify modeling under varying \textit{adversarial conditions}.

\bhd{Procedure Modeling}
Our modeling focuses on procedural flexibility in QKD standards, which often leave low-level operational details to implementation, and we capture the resulting ambiguities through two key procedural choices: the \textit{timing of basis exchange} and the \textit{directionality of information reconciliation}. For basis exchange after qubit transmission, we consider two common configurations: (a) Alice announces her basis before Bob ({\seqa}) and (b) Bob announces his measurement basis before Alice ({\seqb}). For reconciliation, we distinguish \textit{forward} error correction (FEC), where Alice sends parity information to Bob, \textit{reverse} reconciliation (REC), where Bob sends correction data to Alice, and \textit{two-way} interactive reconciliation (TWEC).

\bhd{Session Management}
To keep analysis tractable, we bound the model to a single QKD session as sequential sessions admit unbounded transition depth through recursive measurement calls and internal transition loops. We still support an unbounded number of protocol instances, i.e., instances that may differ in timing, message delivery, and reconciliation order, and we enforce quantum properties in them by restricting state transitions to one unique \textit{variable} of an \textit{encoded qubit} or an \textit{entangled pair} in the state memory.

\subsubsection{Quantum Symbolic Model}
\label{subsec:quant_sym_model}
Probabilistic quantum operations resist direct encoding in labeled transition systems, since verification becomes undecidable~\cite{ying2019model}. The core difficulty is modeling quantum operations whose outcomes govern classical post-processing decisions, e.g., ETSI GS QKD 005 V1.1.1\cite{etsi_qkd_005} describes \textit{generated key} as:

\begin{quote}
\textit{``The security statement of a QKD protocol is of a probabilistic nature. The final key can be claimed to be completely random and completely private, except with a probability $\epsilon$.''}
\end{quote}

\noindent
We represent quantum operations as symbolic functions: uninterpreted terms whose algebraic properties encode protocol behavior without assigning probability amplitudes or Hilbert-space semantics. This abstraction preserves the causal dependencies between quantum measurements and classical post-processing steps, which is sufficient to detect the three vulnerability classes we report. It does not capture coherent attacks, superposition-based strategies, or channel noise. Attacks that exploit those properties require a probabilistic model and fall outside the scope of this analysis. We capture protocol behavior through \textit{logical conditions} that constrain how symbolic functions are manipulated in system transitions. This eliminates the need for non-deterministic transitions, i.e., a unique state for all possible outcomes. These symbolic functions thereby establish the bridge: outcomes of quantum measurements determine the inputs of classical messages.

Consider the example of \textit{parameter estimation phase}. We represent Bob's computation of quantum error with a symbolic function \(NoEve_B=check(s_A, s_B)\) that encodes whether he detects the presence of an eavesdropper \(Eve\). We define \(check(s_A, s_B)\) using a logical condition that verifies whether Bob's measured qubit \(q_x\) and received classical messages (\(s_A\) and \(b_A\)) are consistent with the qubit \(q_A\) originally prepared by Alice:
\[check(s_A, s_B):=~(q_x = q_A) \land s_A = f(q_x, b_A, b_B)\]

\noindent
This logical condition controls a binary transition from \textit{parameter estimation phase}:
\[\texttt{next\_state} :=
\begin{cases}
{KeyGeneration} & \text{if } check(s_A, s_B) \\
{Termination} & \text{otherwise}
\end{cases}\]

\textit{Logical conditions} over symbolic functions also allow us to abstractly model the \textbf{adversarial behavior}, while excluding the computational mechanisms (e.g., quantum coherent attacks, brute-force attacks).
Consider a scenario when Bob detects the presence of an eavesdropper \(Eve\), i.e., $check(s_A, s_B) = false$, and  \textit{parameter estimation} proceeds to $Termination$. We model it as the logical result of a measure-resend attack performed by \(Eve\) such that she intercepts and measures \(measure(q_A, b_{Eve})\) an encoded qubit \(q_A:=encode(k_A^{raw},b_A)\) transmitted over the quantum channel.

Instead of modeling concrete values for this measurement, which would require a different state for each possible outcome, we define two logical outcomes of this measurement: (i) \(Eve\) knows preparation basis \(b_A\) used to create \(q_A\) in which case \(measure(q_A, b_{Eve})\) reduces to \(k_A^{raw}\) and \textit{learns} the raw key \(k_A^{raw}\). Alternatively, (ii) \(Eve\) generates a \textit{fresh} measurement basis \(b_{Eve}\) to measure \(q_A\), which makes \(measure(q_A, b_{Eve})\) irreducible in subsequent states. We define these outcomes with the conditional equation:
\[measure(q_A, b_{Eve}) := k_A^{raw} \Leftrightarrow EveKnows(b_A)~\land~b_{Eve}=b_A\]

\subsection{Adversary Modeling}
\label{subsection:adv_model}
We analyze each protocol model \(\mathcal{P} \in \{P_{EB}, P_{PM}\}\) under the network adversary \(Eve^+\), yielding instantiated models \(\mathcal{P}^{Eve^+}\).

\bhd{MiTM on the Network -- \(Eve^+\)}
The \(Eve^+\) profile is fixed in \S\ref{sec:threat_model}. Here we only encode it as a \textit{non-deterministic transition system} over channels:
\[Eve^+ = Eve~||~Adv^C \]
For the quantum channel between \(\mathcal{L}_A\) and  \(\mathcal{L}_B\), \(Eve^+\) can allow a qubit \(q_A\) to pass undisturbed \(forward(q_A)\), or intercept it \(store(q_A)\) for measurement, or forward it at a later time (delay). In our setup, the decision to intercept or forward the qubit is made by an environmental variable \texttt{adv\_action} whose value is \textit{non-deterministically} chosen by \(Eve^+\). The non-deterministic nature of both the channels and the adversary is essential for automated reasoning about feasible adversarial strategies achievable through combined manipulation of quantum and classical actions.

To enforce quantum constraints on \(Eve^+\), we model symbolic functions with linear (non-persistent) states. As a result, \(Eve^+\) cannot duplicate or reuse qubits (recall no-cloning theorem). For instance, if \(Eve^+\) intercepts qubit \(q_A\), i.e., \(store(q_A)\), she can choose to \(measure(q_A,b_{Eve})\) it or \(forward(q_A)\) it over a quantum channel. Either action \textit{deletes} \(q_A\) from \(Eve^+\)'s knowledge, replacing it with:
\[q_A :=
\begin{cases}
{measure(q_A,b_{Eve})} & \text{if } measured \\
{null} & \text{if } forwarded
\end{cases}\]

\subsection{Security Properties Modeling}
\label{subsubsection:security_properties}
We require that the \textit{key secrecy} and \textit{entity authentication} guarantees of QKD remain valid when the generated key is used in any classical cryptographic primitive~\cite{portmann2014cryptographicsecurityquantumkey}. To this end, we redefine the security goals of QKD in terms of classical security properties~\cite{mosca2013quantum}.
 We model these security goals as \textit{lemmas}, defined using logical equations over transition labels. Each transition label annotates states that occur during all possible execution paths of the protocol. For example, \(Commit_A(B,sid,k_{AB})\) represents a state where \(A\) successfully generates a key \(k_{AB}\) for a session identified by \(sid\).

\bhd{Session Key Secrecy}
\(Secret\_x(k_{xy})\) marks the state in which participant \(x \in \{A,~B\}\) generates \(k_{xy}\) at timestamp \(i\) and \(DetectedEve\) indicates that \(x\) detects eavesdropper \(Eve\) and terminates the protocol at timestamp \(k\). We represent the state where an adversary learns the value of \(k_{xy}\) with \(K(k_{xy}
)\) at some other timestamp \(j\). During analysis, we verify this property against our protocol model, which checks that the generated key \(k_{xy}\) remains secret during and after the communication exchange. A violation of this property implies that an adversary can \textit{learn} $k_{xy}$.  This is formalized as:

\begin{mybox}
\ihd{\textbf{Lemma 1} (Key Secrecy)} A protocol \(P\) satisfies secrecy of participant \(x\)'s key if for every trace \(\alpha \in traces(P)\):

\noindent
\(\forall k_{xy} ~ \#i.  ~  \text{ Secret}_x(k_{xy}) \in \alpha_i \xRightarrow{\hspace{10pt}}\)

\(\hfill\nexists j. \text{ K}(k_{xy})@\in \alpha_j ~ \lor\)
\(\exists k. \text{ DetectedEve}_x \in \alpha_k\)
\end{mybox}

\bhd{Entity Authentication}
The second security property corresponds to entity authentication at the end of a protocol session. Intuitively, when a participant completes a session, resulting in a session key, then the intended peer must have participated in the session. For example, when Alice \(A\) initiates a session with PSK \(pskAB\) and completes it with the key $k_{AB}$, then the intended peer Bob \(B\) must have generated the same key $k_{AB}$ for the same session.

We define this property using Lowe's hierarchy of authentication guarantees~\cite{lowe1997hierarchy}, specifically with the \textit{non-injective} key agreement between \(A\) and \(B\) with two transition labels: \texttt{Commit}, a participant's commitment to a session, and \texttt{Running}, the intent to complete the protocol with a specified user. Non-injective agreement ensures that for any instance of \texttt{Commit}, there exists at least one corresponding instance of \texttt{Running} from the other party with matching parameters. This is formalized for participant \( A \) as follows:

\begin{mybox}
\ihd{\textbf{Lemma 2} (Non-injective Agreement for \(A\))} If \(A\) commits to a session, then \(B\) must have participated and they agree on the key:

\noindent
\(\forall A ~ B ~ \mathit{sid} ~ k_{AB} ~\#i. ~ \text{ Commit}_A(B, \mathit{sid},k_{AB})\in \alpha_i \xRightarrow{\hspace{10pt}}\)

\(\hfill\exists j.\text{ Running}_B(A, \mathit{sid},k_{AB})\in \alpha_j\)
\end{mybox}

\ihd{Executability}
Before verifying security properties, we ensure that the system reaches a valid end state where \(A\) and \(B\) have executed the protocol session to completion for each model. This serves as a sanity check to ensure that security properties are not verified simply because the model is not executable, e.g., due to unreachable states or invalid protocol behavior.
\begin{mybox}
\ihd{\textbf{Lemma 3} (Key Generation Executability)} A protocol \(P\) is executable if both \(A\) and \(B\) generate a symmetric key $k_{AB}$ for at least one trace \(\alpha \in traces(P)\):

\noindent
\(\exists A ~ B ~ \mathit{sid} ~ k_{AB} ~ \#i ~ \#j.\)

\(\hfill\text{Commit}_A(B, \mathit{sid},k_{AB})\in \alpha_i \land \text{Commit}_B(A, \mathit{sid},k_{AB})\in \alpha_j\)
\end{mybox}

\subsection{Verification \& Validation}
We now describe how we perform systematic and comprehensive analysis of models under threat models described in \S\ref{subsection:QKD_model} and \S\ref{subsection:adv_model}. Our \textbf{QVerify}~\cite{QVerify2025} framework is publicly available and provides all models, \texttt{m4} templates, automated scripts, and HTML reports needed to reproduce our Tamarin analysis (adapting scripts from~\cite{basin2021emv}).

\renewcommand{\arraystretch}{1.05}
\begin{table}[t]
\centering
\caption{Verification under $Eve^+$.}
\vspace{-1.5em}
\noindent\parbox{\linewidth}{\scriptsize \center \verified{} verified, \falsified{} falsified.}
\vspace{0.35em}
\label{tab:results_group}
\resizebox{\linewidth}{!}{
\begin{tabular}{@{}clcccccc@{}}
\toprule
\# & Model & Exec & sec
& alive & wAgree & niAgree & Time \\
\midrule
1 & $Eve^+$\_TWEC   & \verified   & \falsified{}$^{1}$ & \verified{}  & \verified{}  & \falsified{}$^3$ & 13m31s \\
2 & $Eve^+$\_REC    & \verified{} & \falsified{}$^{1}$ & \verified{}  & \verified{}  & \verified{}/\falsified{}$^3$ & 2m12s \\
3 & $Eve^+$\_FEC    & \verified{} & \falsified{}$^{1}$ & \verified{}  & \verified{}  & \falsified{}$^3$/\verified{} & 3m58s \\
\bottomrule
\end{tabular}}
\vspace{0.35em}
\noindent\parbox{\linewidth}{\raggedright\textbf{(a)}~\textit{EB-QKD verification under $Eve^+$ for error reconciliation configurations: TWEC, REC, FEC.}}
\vspace{0.6em}

\resizebox{\linewidth}{!}{
\begin{tabular}{@{}clcccccc@{}}
\toprule
\# & Model & Exec & sec
& alive & wAgree & niAgree & Time \\
\midrule
1 & $Eve^+$\_B2A\_TWEC & \verified   & \falsified{}$^{2}$/\verified{} & \falsified{}$^{3}$ & \falsified{}$^{3}$ & \falsified{}$^{3}$ & 7m30s \\
2 & $Eve^+$\_B2A\_REC  & \verified{} & \verified{}  & \verified{}/\falsified{}$^{3}$ & \verified{}/\falsified{}$^{3}$ & \verified{}/\falsified{}$^{3}$ & 59.50s \\
3 & $Eve^+$\_B2A\_FEC  & \verified{} & \falsified{}$^{2}$/\verified{} & \falsified{}$^{3}$/\verified{} & \falsified{}$^{3}$/\verified{} & \falsified{}$^{3}$/\verified{} & 30.38s \\
4 & $Eve^+$\_A2B\_TWEC & \verified{} & \falsified{}$^{2}$ & \falsified{}$^{3}$ & \falsified{}$^{3}$ & \falsified{}$^{3}$ & 7m59s \\
5 & $Eve^+$\_A2B\_REC  & \verified{} & \falsified{}$^{2}$ & \verified{}/\falsified{}$^{3}$ & \verified{}/\falsified{}$^{3}$ & \verified{}/\falsified{}$^{3}$ & 1m31s \\
6 & $Eve^+$\_A2B\_FEC  & \verified{} & \falsified{}$^{2}$ & \falsified{}$^{3}$/\verified{} & \falsified{}$^{3}$/\verified{} & \falsified{}$^{3}$/\verified{} & 45.38s \\
\bottomrule
\end{tabular}}
\vspace{0.35em}
\noindent\parbox{\linewidth}{\raggedright\textbf{(b)}~\textit{PM-QKD verification under $Eve^+$ for configurations basis announcement (\seqa, \seqb) and reconciliation (TWEC, REC, FEC).}}
\vspace{0.35em}
\noindent\parbox{\linewidth}{\scriptsize Superscripts map falsifications to discovered vulnerabilities: \\ $^1$V1 (Subverted Entanglement Injection), $^2$V2 (Basis-Deferred Measurement), $^3$V3 (Message Reflection). See Section~\ref{section:traceanalysis}.}
\vspace{-14pt}
\end{table}

\bhd{Model Templates using \texttt{m4}}
Multiple procedures, basis orders, and reconciliation modes would otherwise require many hand-maintained Tamarin files. We use the \texttt{m4} macro processor on one template: flags select PM versus EB, \seqa{} versus \seqb{}, and FEC versus REC versus TWEC.

Instantiations yield nine \textit{target models} under \(Eve^+\) (Table~\ref{tab:results_group}), with preprocessor switches selecting which lemmas run so that falsified results map cleanly to a configuration, as documented in the \textbf{QVerify} repository~\cite{QVerify2025}.
Across all nine configurations, the generated target models range in complexity from 26 to 34 rewriting rules.

\bhd{Automated Verification}
Given a protocol variant, threat model, and security property, we generate a \textit{target model} from the template and analyze it with the Tamarin Prover, using scripts adapted from~\cite{basin2021emv} to summarize proofs in HTML. Scripts in \textbf{QVerify} automate instantiation, batch verification, and compilation of outcomes into tables aligned with Table~\ref{tab:results_group}. See~\cite{QVerify2025}. Secrecy lemmas fail when classical leakage combines with quantum correlations, and authentication lemmas fail when classical commitments are not bound to the same session's quantum exchange.

\bhd{Counterexample Analysis}
When Tamarin reports \(falsified\), we inspect the HTML summary, then use interactive mode on failing target models to obtain traces, attribute violations to procedure or adversary choices, and refine the template. During development, \textit{executability} lemmas (Lemma~3) serve as sanity checks so failures are not artifacts of dead models.

\section{Critical Security Gaps in QKD Standards}
\label{section:traceanalysis}

Verification under adversary \(Eve^+\) exposes three classes of specification-level vulnerabilities (timing, ordering, and role direction) that break secrecy or authentication across the nine configurations we analyze (Table~\ref{tab:results_group}). Each class shows how \(Eve^+\) exploits missing control-plane constraints in the modeled procedures to defeat the checked symbolic properties. For V1 the root cause is a classical timing and synchronization omission. The quantum-side re-encode maneuver is the exploit path enabled by that omission.

\subsection{\texttt{V1}: Subverted Entanglement Injection}
\label{attack:v1}
\bhd{Overview}
EB QKD protocols lack a mechanism to verify that communicating parties receive qubits at consistent timestamps. Synchronization depends entirely on classical-channel control signals susceptible to time-shift manipulation~\cite{qi2005time, zhao2008quantum}. This gap creates a causal link between the timing of protocol operations and the secrecy of the generated key. An adversary \(Eve^+\) who introduces a clock offset between Alice and Bob can force one participant to announce measurement bases before the other has initiated the session. The premature basis disclosure exposes private measurement outcomes from the quantum phase that \(Eve^+\) uses to derive the session key.

\begin{figure}[t]
    \centering
    \includegraphics[width=\linewidth]{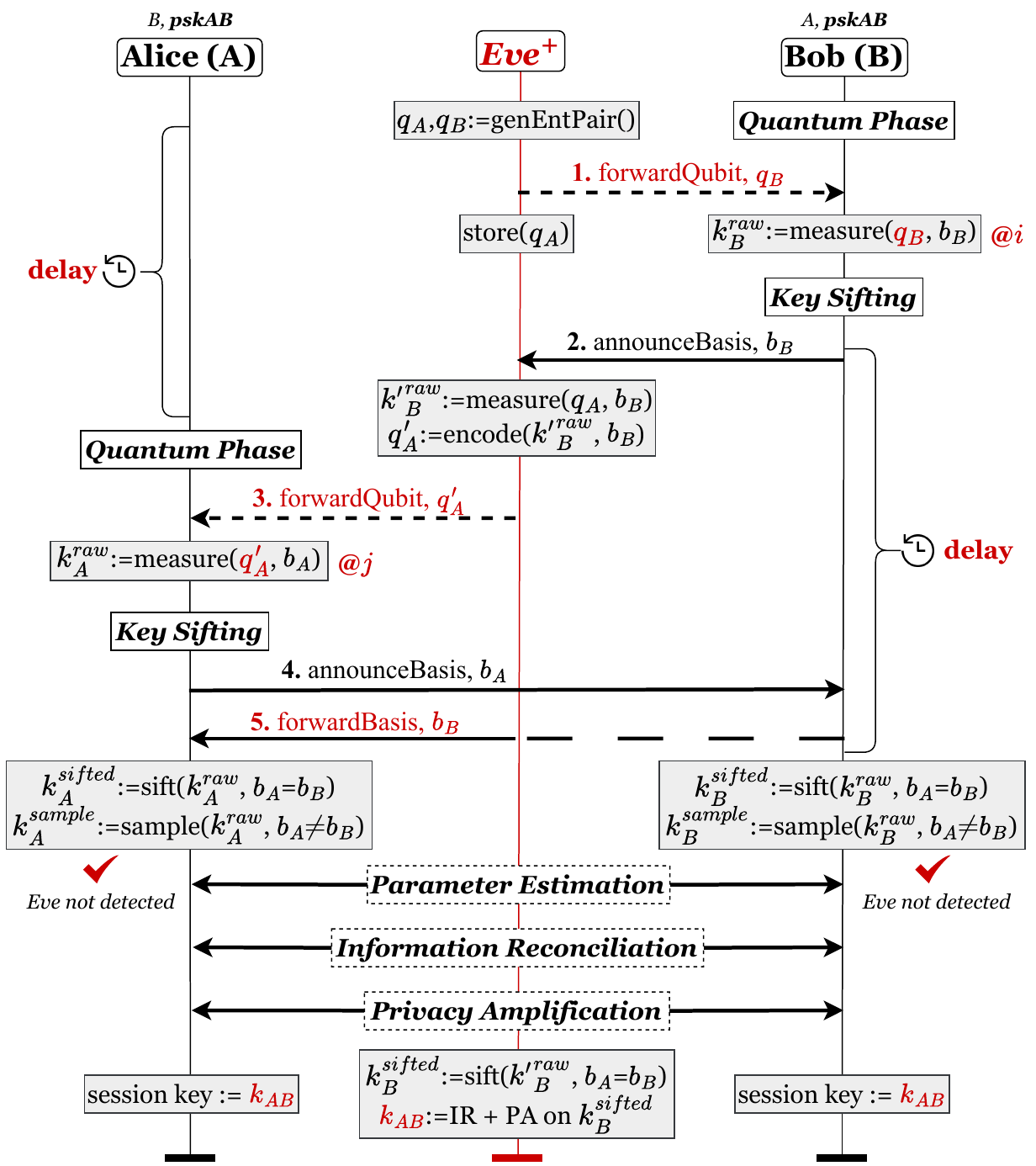}
    \caption{Subverted Entanglement Injection attack. \(Eve^+\) controls the entanglement source to retain one qubit pair while the other is sent to one of the parties, e.g., Bob. She uses Bob's basis announcement to measure her retained qubits, reconstructs the anti-correlated key, and re-encodes a qubit stream for Alice. The information exchanged during the classical post-processing enables final key to be recovered undetected.}
    \label{fig:attack_A}
    \vspace{-5pt}
\end{figure}

\bhd{Detection}
The attack surface became apparent when \(P_{EB}\) falsified \textit{\textbf{Lemma 1} (Key Secrecy)} under the threat capabilities of \({Eve^+}\)
(see Table~\ref{tab:results_group}(a)). Analysis of the counterexamples reveals a scenario where \(Eve^+\) creates a time shift between participants (Alice and Bob). This causes Alice and Bob to initiate a QKD session \(sid\) with qubits received at different \textit{timestamp} \(j\) and \(i\), respectively, with a time shift \(i>j\) or \(j>i\) (see Figure ~\ref{fig:attack_A}). The time shift enables \(Eve^+\) to trigger one participant (e.g., Bob) to announce their measurement basis for \(sid\) before the other participant (e.g., Alice) has initiated \(sid\). As a result, \(Eve^+\) \textit{learns} the private measurement results of one participant during the quantum phase, which are used to \textit{derive} the session key in later protocol steps.

\bhd{Attack Procedure}
Figure~\ref{fig:attack_A} shows the attack when \(Eve^+\) opens the session with Bob.
\begin{enumerate}
    \item \textit{Pair splitting and basis-driven measurement:} \(Eve^+\) keeps one branch \(q_A\) of each entangled pair, sends \(q_B\) to Bob \circled{1}, then reads Bob's basis announcement \(b_B\) \circled{2} and measures \(q_A\) in matching bases. Entanglement correlations ensure that her raw string \(k_{B}^{'raw}\) agrees with Bob's \(k_{B}^{raw}\) up to an error rate bounded by the channel QBER. Under our Aer setup, sampled CHSH statistics remain compatible with acceptance (\S\ref{subsec:simulated_results})\footnote{Entangled qubits are correlated such that the measurement outcome of one qubit instantaneously determines the outcome of the other depending on their entangled state (e.g., Bell states like \( |\Phi^+\rangle \) exhibit perfect Z-basis correlation, and anti-correlated states such as \( |\Psi^-\rangle \) are correlated in the X basis and anti-correlated in the Z basis).
    }.

    \item \textit{Forged forward line and sifting:} She prepares \(q'_A\) from \(k_{B}^{'raw}\) and \(b_B\), sends it to Alice \circled{3}, and relays basis traffic so Alice and Bob sift and sample as usual \circled{4}--\circled{5}.

    \item \textit{Parameter estimation and reconciliation:} As the sampled strings are statistically similar, CHSH checks can pass under our Aer setup (we report the CHSH comparison with Qiskit in \S\ref{subsec:simulated_results}). Alice and Bob run reconciliation and privacy amplification to \(k_{AB}\).

    \item \textit{Key copy:} \(Eve^+\) applies the same public post-processing to \(k_{B}^{'raw}\) and learns \(k_{AB}\).
\end{enumerate}

\bhd{Root Cause}
EB QKD protocols do not have native mechanisms to verify qubit timestamps and ensure synchronization without relying on external sources~\cite{etsi_qkd_005, ekert1991quantum, ent_middle1}. Instead, they rely on control signals over the classical channel (e.g., time synchronization protocols like NTP~\cite{itu_y3802,etsi_qkd_011} and analog sync channels~\cite{etsi_qkd_012,itu_y3802,itu_y3803}) that are known to be vulnerable to time-shift attacks~\cite{mahlous2024quantitative, perry2021devil,li2023secure,qi2005time, zhao2008quantum}. An adversary can launch these attacks to introduce a clock offset between the communicating participants (i.e., Alice and Bob) to cause one participant to announce basis earlier than the other, which compromises the secrecy of the session key.

\bhd{Impact}
Within the scope of our model, the attack yields the session key \(k_{AB}\) for the compromised session, enabling decryption and injection of encrypted traffic between Alice and Bob. Whether this affects subsequent sessions depends on deployment-specific key management practices, specifically whether \(k_{AB}\) is used to derive or refresh PSKs. We do not model key management across sessions.

\subsection{\texttt{V2}: Basis-Deferred Measurement}
\label{attack:v2}

\bhd{Overview}
We observe that PM QKD specifications do not define explicit ordering (i.e., the true sequence of message exchange) for the \textit{measurement basis announcement} phase~\cite{itu_FG_QIT4N_D2_3_1,etsi_qkd_005}. This creates two possible execution sequences, which we model in \S\ref{subsec:protocol_model} as: {\seqa}, where basis disclosure may occur before qubit measurement, and {\seqb}, where it is deferred until afterward. Our analysis reveals that PM protocols conforming to {\seqa} are inherently less secure. By announcing Alice's basis before Bob measures, the {\seqa} ordering allows \(Eve^+\) to hold stored qubits and measure them in the correct basis after the announcement. Because \(Eve^+\) measures before forwarding rather than attempting to copy the qubit, the no-cloning theorem is not violated. Instead, \(Eve^+\) re-encodes a fresh qubit from the now-known basis and forwards it, so parameter estimation sees no error increase. The protocol's security therefore depends on ordering assumptions that the standard does not enforce.

\bhd{Detection}
We obtain the \textit{basis-deferred measurement attack} after observing our models \(P_{PM}\) violate the \textit{\textbf{Lemma 1} (Key Secrecy)} under the threat capabilities of \({Eve^+}\) as defined in \S\ref{sec:threat_model} under the {\seqa} configurations (shown in Table~\ref{tab:results_group}(b)). Analysis of the counterexamples reveals a scenario in which Alice announces her measurement basis before Bob completes qubit detection. Positioned between the participants, \(Eve^+\) intercepts the encoded qubits and measures them in Alice's declared basis to reveal her raw key. She then reconstructs and forwards a copy of Alice's \textit{encoded qubits} to Bob. The reconstructed qubits are consistent with Alice's raw key and announced basis, the protocol proceeds to \textit{key generation} without error, while \(Eve^+\) derives the final shared key.

\begin{figure}[t]
    \centering
    \includegraphics[width=\linewidth]{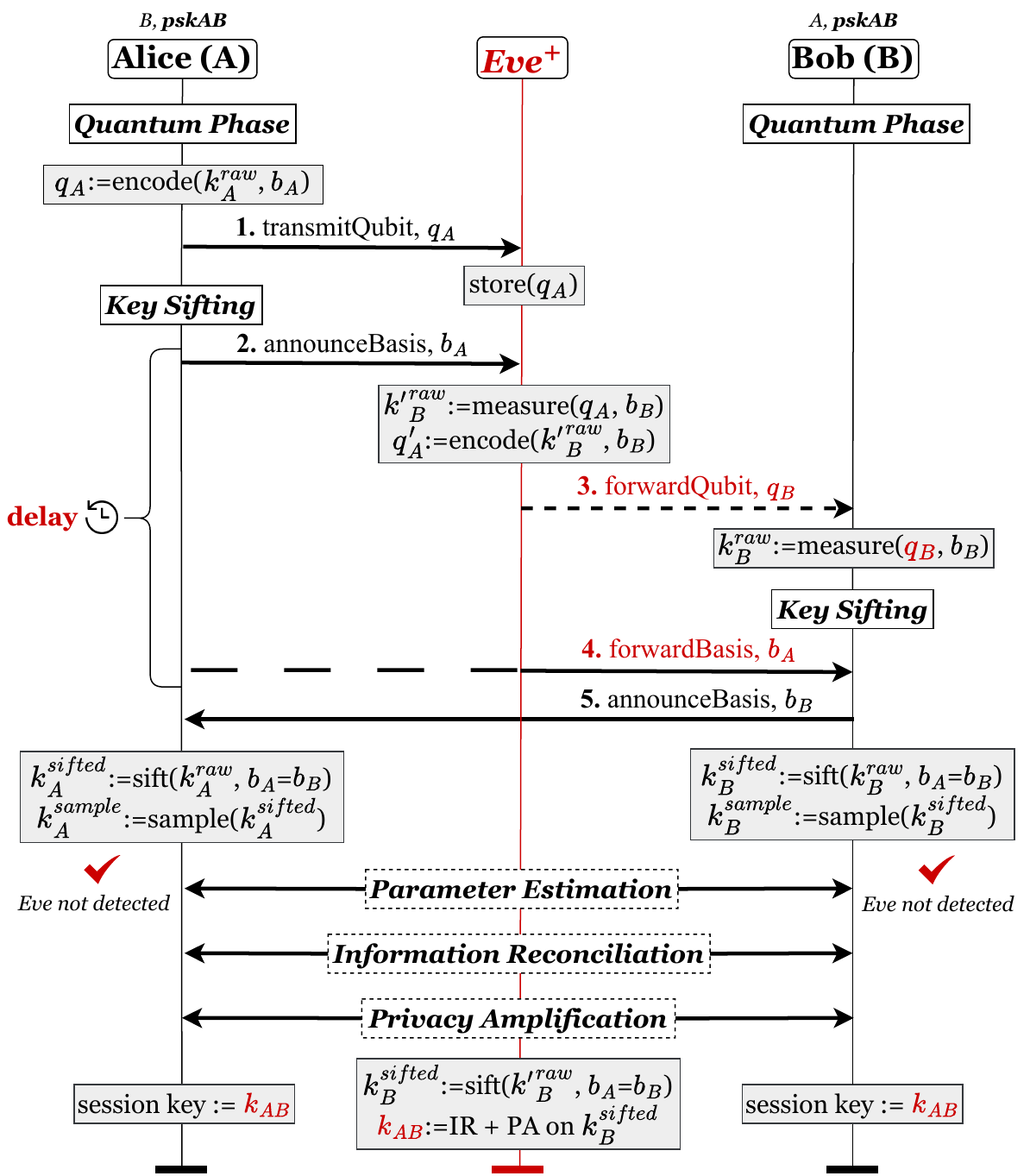}
    \caption{Basis-Deferred Measurement attack. \(Eve^+\) intercepts and stores Alice's qubit stream, then measures it using Alice's publicly revealed basis. She reconstructs and forwards a forged qubit sequence to Bob. With basis forwarding, \(Eve^+\) learns the final key without introducing detectable errors.}
    \label{fig:attack_C}
    \vspace{-10pt}
\end{figure}

\bhd{Attack Procedure}
Figure~\ref{fig:attack_C} shows {\seqa}: \(Eve^+\) buffers \(q_A\) \circled{1} until Alice publishes \(b_A\) \circled{2}\footnote{Storing qubits does not reveal any information to \(Eve^+\). It is retained in memory until \(Eve^+\) decides to either measure or forward the qubit.}.
\begin{enumerate}
\item \textit{Measure after announcement:} \(Eve^+\) measures the buffer in \(b_A\), learns \(k_{A}^{raw}\), and sends a forged \(q'_A\) to Bob \circled{3}. Bob measures with \(b_B\) to obtain \(k_{B}^{raw}\).

\item \textit{Sifting and estimation:} \(Eve^+\) relays bases \circled{4}--\circled{5}. Because \(q'_A\) encodes the same \((k_{A}^{raw},b_A)\) Alice used, sifted and sampled strings align up to noise, parameter estimation passes, and the parties derive \(k_{AB}\).

\item \textit{Key copy:} \(Eve^+\) runs the same reconciliation and privacy amplification on \(k_{A}^{raw}\) and learns \(k_{AB}\).
\end{enumerate}

\bhd{Root Cause}
Neither ETSI GS QKD~\cite{etsi_qkd_005} nor FG QIT4N D2.3.1~\cite{itu_FG_QIT4N_D2_3_1} mandates a sequencing constraint between Alice's basis disclosure and Bob's qubit measurement, leaving this ordering to implementation choice. Implementations that follow the {\seqa} configuration allow Alice to reveal her basis before Bob measures, without requiring synchronization. An adversary can therefore store intercepted qubits and defer measurement until Alice transitions to the \textit{measurement basis announcement} procedure.

\bhd{Impact}
Within the scope of our single-session model, this attack enables \(Eve^+\) to compromise the session key \(k_{AB}\) and to decrypt or inject encrypted messages between Alice and Bob for that session. We do not model key management across sessions.

\subsection{\texttt{V3}: Message Reflection}
\label{attack:v3}

\bhd{Overview}
This attack shows that without binding messages to peer direction and session roles, \(Eve^+\) can reflect MAC-protected traffic so a party effectively talks to itself while believing it talks to the peer.

\begin{figure}[b]
    \centering
    \includegraphics[width=\linewidth]{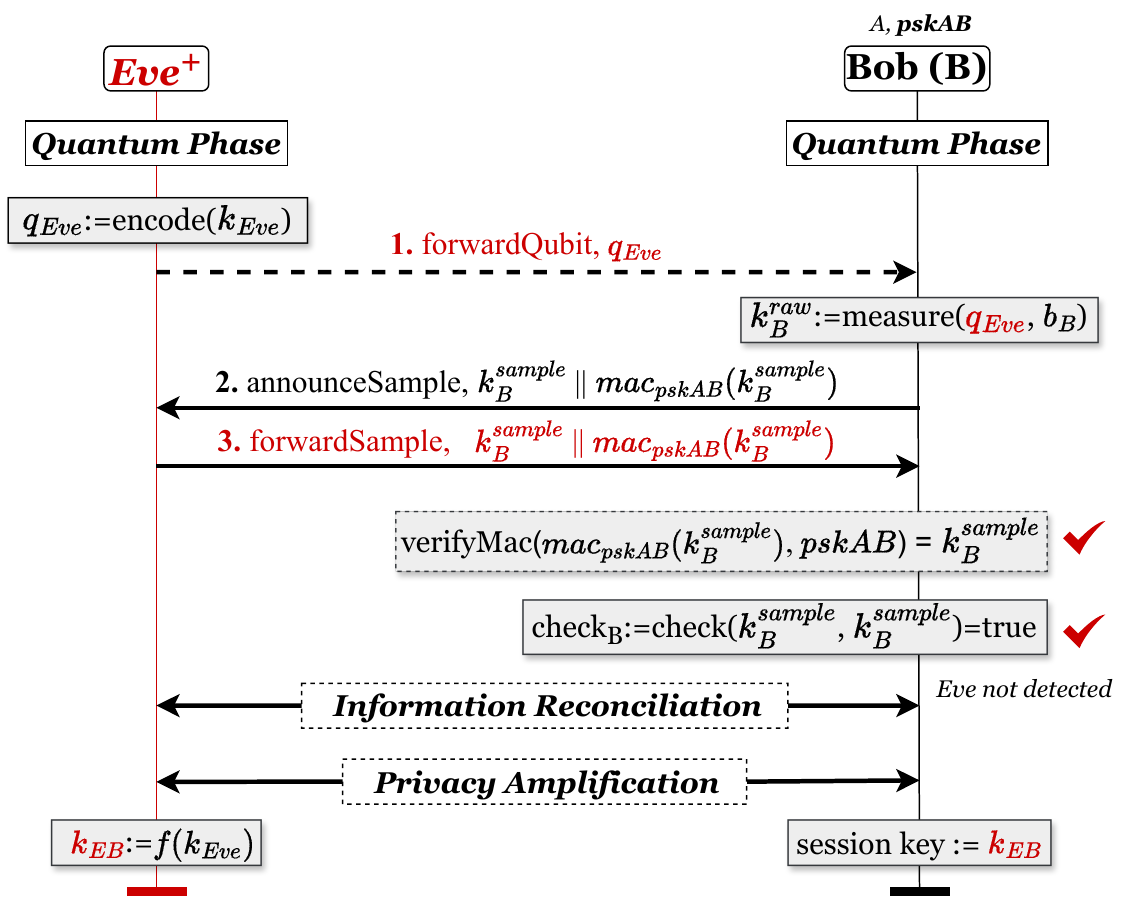}
    \caption{Self-Authentication via Message Reflection. \(Eve^+\) reflects a valid MAC-tagged message back to its originator, who accepts it as authentic. The protocol completes with a shared key based on self-reflected inputs, violating mutual authentication.}
    \label{fig:attack_D}
    \vspace{-15pt}
\end{figure}

\bhd{Detection}
We uncover multiple instances of \textit{reflection attacks} in our analysis of models \(P_{PM}\) and \(P_{EB}\). Our verification results indicate violation of all the agreement properties \textit{\textbf{Lemma 2} (Aliveness, Weak, and Non-injective Agreement)} and in some cases, \textit{\textbf{Lemma 1}(Key Secrecy)} (see Table~\ref{tab:results_group}(a) and (b)). The counterexamples yield protocol states in which a participant completes a session with itself, i.e., \textit{Alice-talks-to-Alice} or \textit{Bob-talks-to-Bob}, while believing they are talking to each other. In each case, the MAC verification check \(verify(mac_{psk}(msg),psk)=msg\) succeeds because identities are not properly bound to the MAC tags, and it does not check if a message is reflected. Consequently, \(Eve^+\) can force a participant (e.g., Bob) to
generate a session key from adversary-generated qubits using reflected messages (see Figure~\ref{fig:attack_D}).

Figure~\ref{fig:attack_D} traces a concrete instance during parameter estimation.
\begin{enumerate}
\item \(Eve^+\) drives Bob with adversarial qubits \(q_{Eve}\) \circled{1}.

\item Bob publishes a sampled string \(k_{B}^{sampled}\) \circled{2}. \(Eve^+\) reflects it unchanged \circled{3}. The MAC still verifies under \(pskAB\), so Bob accepts his own message as if it came from Alice.

\item \(check(k_{B}^{sampled},k_{B}^{sampled})\) is trivially zero error, estimation succeeds, and Bob completes key generation from Alice-independent data.
\end{enumerate}

\bhd{Root Cause}
QKD standards~\cite{itu_x1712,itu_y3803,etsi_qkd_005} mandate MAC-based authentication for classical post-processing messages but do not specify that role metadata, session identifiers, or directionality must appear in the MAC input. Without these bindings, a reflected message verifies correctly under the shared PSK. The omission permits role misbinding and agreement on invalid cryptographic keys within the modeled session.

\bhd{Impact}
Within the modeled session, the adversary deceives a participant into accepting its own reflected messages as if they were sent by the peer. The adversary can thereby bypass entity authentication in that session and, in some cases, induce key generation from self-reflected inputs. The attack violates the checked agreement properties under our symbolic model. It does not by itself establish impersonation across later sessions, which our model does not cover.

\subsection{Simulated Results}
\label{subsec:simulated_results}

\bhd{Hybrid Exploit Path in \texttt{V1}}
\texttt{V1} is a specification-level vulnerability: a control-plane timing and synchronization omission that adversary \(Eve^+\) exploits via a hybrid quantum re-encode attack. In the counterexample trace for subverted entanglement injection, \(Eve^{+}\) uses that classical timing gap to perform a quantum-side maneuver: she measures her share of an entangled bipartite system before the honest peer performs their measurement. Using the observed outcomes, \(Eve^{+}\) then actively prepares and forwards new qubits. Unaware of the quantum manipulation, the remote party proceeds with sifting and parameter estimation as usual.

This attack hinges on the adversary's ability to encode new quantum states based on the measurement results obtained from the entanglement-side. Essentially, measurement outcomes from one branch are mapped, or "forged," to produce valid quantum states for the other branch. To assess whether this symbolic attack remains compatible with observables an honest entanglement-based participant would accept under our Aer setup, we simulate the corresponding scenario in Qiskit~\cite{ibm_qiskit} and compute error rates and related statistics.

\bhd{Error distributions in Qiskit} To assess the detectability of the \texttt{V1} adversarial maneuver at the quantum level, we compare sampled error distributions, the primary tool for adversary detection, under honest entanglement-based (EB) operation and under adversarial models. We simulate circuits in Qiskit Aer with \(240\) Bell pairs per circuit, \(120\) shots per run, and four independent random-basis trials of Alice and Bob measurement settings. For each trial we aggregate relative CHSH error into \(5\%\) buckets of \(\lvert (S-S_{\mathrm{ideal}})/S_{\mathrm{ideal}}\rvert\) with \(S_{\mathrm{ideal}}=-2\sqrt{2}\), then average and renormalize the histograms across trials. The honest EB baseline uses a depolarizing noise model on single-qubit gates (\(0.0012\)), two-qubit \(\mathrm{CX}\) gates (\(0.006\)), and symmetric readout error (\(0.008\)), while the attack circuits are noiseless in our stock configuration.

\begin{figure}[t]
    \centering
    \subfigure[Standard Measure-Resend Attack.]{
        \includegraphics[width=0.85\linewidth]{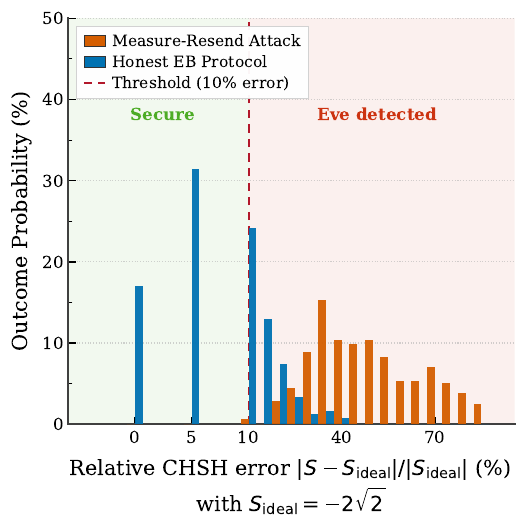}
        \label{fig:distCHSH}
    }
    \vspace{-6pt}
    \subfigure[V1 Entanglement-Injection Attack.]{
        \includegraphics[width=0.85\linewidth]{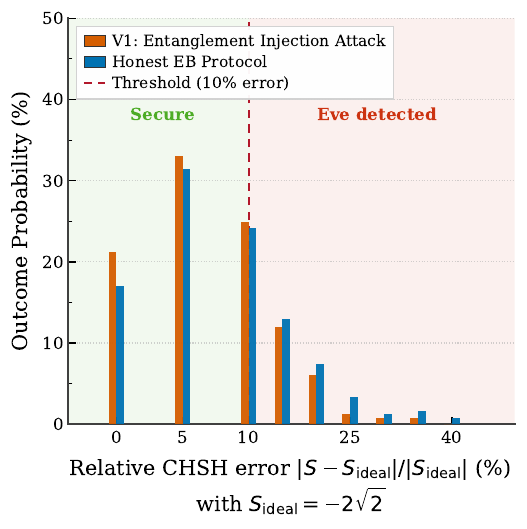}
        \label{fig:distCHSH_V1}
    }
    \caption{Relative CHSH violation error for adversarial and honest entanglement-based (EB) runs. Each panel uses \(S_{\mathrm{ideal}}=-2\sqrt{2}\), \(5\%\) bins of relative error \(\lvert S-S_{\mathrm{ideal}}\rvert/\lvert S_{\mathrm{ideal}}\rvert\), outcome probability (\%) on the vertical axis, a \(10\%\) relative-error threshold (dashed), and orange for the attack versus blue for honest EB.}
    \label{fig:relerr_chsh_panels}

\end{figure}

Figure~\ref{fig:relerr_chsh_panels} presents the results, highlighting a \(10\%\) relative-error threshold (dashed). Figure~\ref{fig:distCHSH} displays the comparison between a standard measure-resend attack and honest EB operation, while Figure~\ref{fig:distCHSH_V1} contrasts the \texttt{V1} entanglement-injection attack with honest EB as described in Section~\ref{attack:v1}. In both cases, the simulated error distributions for honest and adversarial scenarios are statistically indistinguishable across all trials. Under our Aer configuration, conventional CHSH-based parameter estimation would therefore not flag the compromised session described in Section~\ref{attack:v1} when adversarial alignment of forwarded quantum states preserves the sampled statistics. These runs provide quantum-level evidence compatible with the symbolic trace rather than a deployment-wide detectability claim~\cite{QVerify2025}.

\bhd{Classical Control-Plane Vulnerabilities in \texttt{V2} and \texttt{V3}} The vulnerabilities underlying \texttt{V2} and \texttt{V3} arise from \textit{classical} control-plane omissions that adversary \(Eve^+\) can exploit. In \texttt{V2}, basis disclosure is ambiguously ordered relative to measurement. In \texttt{V3}, MAC-verified messages carry no role or direction binding.
For both variants, the Tamarin counterexamples require no quantum-circuit layer to account for trace behavior, because each exploit operates entirely within classical sequencing, authentication structure, and the modeled message algebra.

\vspace{-4pt}
\section{Proposed Countermeasures}
\label{sec:countermeasures}

We propose protocol-level countermeasures that address the vulnerabilities in QKD protocols we identify under \(Eve^+\). Each countermeasure targets specific failure modes while maintaining compatibility with current QKD systems.

\subsection{Timing and Authentication Binding (CM1--CM2)}

Two classes of vulnerability share a common root: the absence of explicit sequencing constraints (\texttt{V1}, \texttt{V2}) and the absence of role and identity binding in MAC inputs (\texttt{V3}). We address both with coordinated specification additions.

\textbf{CM1: Measurement commitment phase.}
QKD protocols permit basis announcement before qubit measurement, creating a timing window an adversary can exploit. We introduce an explicit \textit{measurement commitment phase}: after measuring qubits, both participants transmit a \textit{MAC-protected} commitment before announcing bases, eliminating the window for deferred measurement and timing-based relay.

\textbf{CM2: Identity-bound MAC.}
Although QKD standards mandate MAC-based authentication, they do not define explicit binding of peer identities and roles in MAC inputs. We strengthen authentication by requiring MAC computations to include explicit sender and receiver identities, role metadata, and session context. Formally:
\vspace{-2pt}
\[
\texttt{mac}_{pskAB}({Sender}_{id}, {Receiver}_{id},{session}_t, data)
\]
\vspace{-2pt}
This binding ensures that any deviation in expected peer identity or session role renders the MAC invalid. It prevents reuse of messages across sessions and enforces correct identity resolution in authenticated exchanges, effectively fixing \texttt{V3} (Message Reflection).

Together, CM1 and CM2 eliminate the timing windows and MAC re-use scenarios that enable \texttt{V1}, \texttt{V2}, and \texttt{V3} without modifying quantum subroutines.

\subsection{Verification of Countermeasures}
We extended the Tamarin template with CM1 and CM2 and re-ran the same lemma suite on all nine \(Eve^+\) target models in Table~\ref{tab:results_group}. CM1 closes the \texttt{V1}/\texttt{V2} timing and basis-order windows on every previously failing EB and PM row. CM2 stops \texttt{V3} because reflected tags cannot satisfy the identity-bound MAC inputs. With CM1 and CM2 together, all nine target models verify secrecy and agreement lemmas without changing the quantum subroutines themselves.

\section{Discussion, Disclosure, and Impact}
\label{sec:discussion_section}

\subsection{Scope and Limitations}

Our Tamarin models follow ETSI and ITU-T procedure descriptions and abstract away device-specific timing, detector defects, and implementation-level bugs.
The vulnerabilities we report are therefore protocol-structure issues, not vendor-specific flaws.
A natural question is whether symbolic verification misses attacks that exploit quantum coherence or superposition-based strategies.
Our secrecy and agreement lemmas provide formal evidence against adversary \(Eve^+\) within the symbolic abstraction. They do not certify resistance to coherent or superposition-based strategies, which lie outside that model.
The control-plane omissions that enable V1--V3---timing, ordering, identity binding, and role metadata---remain in the procedure text even when such quantum strategies are in scope, because no coherent attack can supply a binding the standard never required.

Realistic quantum bit-error rates (QBER) affect error-correction thresholds and key-rate estimates, yet QBER does not influence the timing, ordering, and identity-binding omissions that V1--V3 exploit.
These vulnerabilities stem from specification-level omissions rather than noise-tolerance parameters, so a higher or lower error rate leaves them intact.
Key management across sessions introduces additional attack surface that our single-session model does not capture. We defer multi-session analysis to future work.
The single-session scope is consistent with the boundary drawn by ETSI GS QKD 005~\cite{etsi_qkd_005}.

\subsection{Disclosure and Stakeholder Outcomes}
\label{subsec:liaison}
Between March and May 2025 we shared results with ITU-T contributors (including work linked to ITU-T Y.QKDN-TSNfr) and the ETSI ISG-QKD community. ITU-T acknowledged gaps and pointed us to ongoing ETSI formal-methods activity. Participants in the liaison process reported that some commercial deployments add authentication mechanisms beyond the published standard text. Whether those mechanisms overcome the vulnerabilities described in this paper has not been evaluated at this time.

\section{Related Work}
This section groups prior QKD analyses by the technical gap each leaves open.

\bhd{Formal Verification of QKD Protocols}
Symbolic abstractions for quantum protocols~\cite{hirschi2019symbolic} and automated analysis tools~\cite{nagarajan2005automated, nagarajan2002formal} introduced foundational methods but were computationally intensive and restricted to isolated quantum properties. Verification efforts on ETSI protocols~\cite{prevost2024formal} and free-space QKD systems~\cite{fernandez2011formal} addressed specific deployments without modeling the full classical post-processing surface. The gap these works leave is coverage: none jointly models PM and EB variants with authenticated classical channels under one executable template. Our Tamarin models close this gap by encoding both protocol families with \(Eve^+\) under a single template.

\bhd{Security Proofs for QKD Protocols.} Existing comprehensive security proofs~\cite{tomamichel2017largely, kubota2011formal} and their extensions~\cite{zhang2021security, su2020simple, tupkary2024security} provide strong information-theoretic guarantees but universally assume an ideal classical infrastructure. Because of this abstraction, they fail to test the procedural choices left underspecified by ITU and ETSI standards, including basis-announcement order, reconciliation direction, and MAC binding. In our work, we encode these practical operational choices as model parameters, demonstrating how their interaction with adversary capabilities can actively falsify established security lemmas.

\bhd{Hybrid Classical Security Models for QKD.} While previous efforts to integrate QKD into classical Authenticated Key Exchange (AKE) frameworks~\cite{bruckner2023end, kon2024quantum, mosca2013quantum} successfully clarified hybrid security goals, they remain limited to theoretical definitions. Because they lack executable models tied to standard variants, evaluations have been restricted to manual analyses of simplified protocols. We address this gap by introducing a template-driven Tamarin pipeline. Our approach enables scalable, automated verification across nine $Eve^+$ configurations, natively incorporating practical parameters such as reconciliation and basis-order.

\bhd{Attacks on QKD Protocols}
SQKD analyses~\cite{ye2023security}, phase-randomization flaws~\cite{curras2023security}, side channels~\cite{jiang2024side}, and sifting studies~\cite{pfister2016sifting} target hardware imperfections or restricted threat models. These attacks assume device-level flaws and do not examine specification-level ambiguities that persist when devices are ideal. We target protocol-structure vulnerabilities and tie each attack to a concrete lemma failure in Table~\ref{tab:results_group}.

\section{Conclusion and Future Work}
We built symbolic Tamarin models of PM and EB QKD from ETSI and ITU-T sources, instantiated nine configurations under adversary \(Eve^+\), and ran four security checks per protocol configuration for secrecy and agreement. The analysis isolates three specification-level vulnerability classes rooted in classical control-plane omissions: subverted entanglement injection (V1), basis-deferred measurement (V2), and message reflection (V3) (Section~\ref{section:traceanalysis}). Two protocol-level countermeasures restore all checked secrecy and agreement lemmas across the nine configurations under \(Eve^+\) (Section~\ref{sec:countermeasures}): measurement commitment (CM1) and identity-bound MAC (CM2). The key takeaway is that classical control-plane constraints matter alongside the quantum channel: ordering, authentication binding, and timing assumptions require the same scrutiny as optical-layer guarantees when adversary \(Eve^+\) is in scope.

We release \textbf{QVerify} to support independent replication and extension~\cite{QVerify2025}. Future work includes extending the models to multi-session key management and experimental validation on open testbeds when they expose compatible interfaces.

\appendices

\section{QKD Specification Documents}
\label{appendix:tab}

To formalize the abstract model for QKD protocols, we refer to procedures specified in 8 recommendations, 11 specifications, and 4 technical reports published by ITU and ETSI (see Table~\ref{tab:qkd_standards} for the full list), covering both \textit{prepare-and-measure (PM)} and \textit{entanglement-based (EB)} variants.

\balance

\begin{table}[t]
    \centering
    \scriptsize
    \setlength{\tabcolsep}{1.3pt}
    \renewcommand{\arraystretch}{0.85}
    \resizebox{\columnwidth}{!}{
    \begin{tabular}{@{}p{4.15cm}@{\hspace{1.2pt}}p{3.25cm}@{}}
        \hline
        \textbf{Document} & \textbf{Title} \\
        \hline
        ETSI GS QKD 002 V1.1.1~\cite{etsi_qkd_002} & Use cases \\
        ETSI GR QKD 003 V2.1.1~\cite{etsi_qkd_003} & Components and interfaces \\
        ETSI GS QKD 004 V2.1.1~\cite{etsi_qkd_004} & Application interface \\
        ETSI GS QKD 005 V1.1.1~\cite{etsi_qkd_005} & Security proofs \\
        ETSI GS QKD 008 V1.1.1~\cite{etsi_qkd_008} & Module security spec. \\
        ETSI GS QKD 011 V1.1.1~\cite{etsi_qkd_011} & Optical components \\
        ETSI GS QKD 012 V1.1.1~\cite{etsi_qkd_012} & Deployment parameters \\
        ETSI GS QKD 014 V1.1.1~\cite{etsi_qkd_014} & REST key-delivery API \\
        ETSI GS QKD 015 V2.1.1~\cite{etsi_qkd_015} & SDN control interface \\
        ETSI GS QKD 016 V2.1.1~\cite{etsi_qkd_016} & CC protection profile (PM) \\
        ETSI GS QKD 018 V1.1.1~\cite{etsi_qkd_018} & SDN orchestration \\
        Rec.\ X.1710~\cite{itu_x1710} & QKDN security framework \\
        Rec.\ X.1712~\cite{itu_x1712} & QKDN key-manage. security \\
        Rec.\ X.1714~\cite{itu_x1714} & Key combination and supply \\
        Rec.\ Y.3800~\cite{itu_y3800} & QKDN overview \\
        Rec.\ Y.3801~\cite{itu_y3801} & Functional requirements \\
        Rec.\ Y.3802~\cite{itu_y3802} & Functional architecture \\
        Rec.\ Y.3803~\cite{itu_y3803} & Key management \\
        Rec.\ Y.3804~\cite{itu_y3804} & Control and management \\
        TR XSTR-SEC-QKD~\cite{itu_xstr_sec_qkd} & Security considerations \\
        TR FG QIT4N D2.3.1~\cite{itu_FG_QIT4N_D2_3_1} & Quantum layer protocols \\
        TR FG QIT4N D2.3.2~\cite{itu_FG_QIT4N_D2_3_2} & KM and control layers \\
        TR FG QIT4N D2.5~\cite{itu_FG_QIT4N_D2_5} & Standardization outlook \\
        \hline
    \end{tabular}}
    \vspace{3pt}
    \caption{Referenced ETSI and ITU-T QKD documents.}
    \vspace{-25pt}
    \label{tab:qkd_standards}
\end{table}

\section{Artifact Availability}
\label{appendix:artifact}
\vspace{-5pt}
\textbf{QVerify}~\cite{QVerify2025} packages the Tamarin theories and tooling used for this paper. The layout centers on one \texttt{main.m4} entry point and reusable fragments under \texttt{generic\_models/}: PM and EB role and adversary pieces, shared restrictions, basis-announcement options under \texttt{common/Basis\_order/}, error-correction variants under \texttt{common/ER/}, and lemma modules for executability (Lemma~3), secrecy (Lemma~1), and authentication (Lemma~2). The top-level \texttt{Makefile} runs \texttt{m4} with flags for protocol type, adversary, error-correction mode, basis order, and which lemmas to include, then invokes the bundled Tamarin Prover on the generated theory. Helper scripts under \texttt{tools/} compile all target combinations, run batch proof jobs, clean incomplete artifacts, and call \texttt{collect} to build an HTML summary (\texttt{generated\_results.html}) from proof logs under \texttt{output/}. A Docker workflow writes the same artifacts to \texttt{\_output/} when Tamarin runs inside the container, which is the supported path on non-Linux hosts because the shipped Tamarin binary targets Linux. The bundle includes Maude-related helpers and a packaged Tamarin executable so runs do not depend on a system-wide install.

For interactive inspection of falsified goals or trace structure, the Makefile exposes web-style Tamarin sessions (including Docker-backed access on a configurable local port). The \texttt{filesforanalysis/} folder holds example PDF exports of dependency traces from interactive mode, and its README explains how those figures are produced. The repository README documents \texttt{make} targets for single configuration versus full batch regeneration, Docker-based execution, and the HTML summary for generating Table~\ref{tab:results_group}.

{\tiny
\bibliographystyle{IEEEtran}
\bibliography{citations}
}

\end{document}